\documentclass[11pt]{article}
\usepackage{amsmath}
\usepackage{amsfonts}
\usepackage{stmaryrd}
\usepackage{vaucanson}
\usepackage[english]{babel}
\def\Z{\mathbb{Z}}
\begin{document}
\begin{center}
{\bf\Large  An unexpected application of minimization theory to module 
decompositions}

\bigskip

G\'erard Duchamp, Hatem Hadj Kacem, \'Eric Laugerotte \\
\{gerard.duchamp, hatem.hadj-kacem, eric.laugerotte\}@univ-rouen.fr\\  
 \it LIFAR, Facult\'e des Sciences et des Techniques,\\
\it 76821 Mont-Saint-Aignan Cedex, France.
\end{center}
The first step in  the minimization process of an automaton \((\lambda,\mu,\gamma)\)
taking it's multiplicities in a (commutative or not) field, due to 
Sch\"utzenberger, is the construction of a prefix set \(P\) such that the 
orbit \(\lambda\mu(P)\) of the initial vector be a basis of 
\(\lambda\mu(k\langle\Sigma \rangle)\) 
(this amounts to construct a covering tree)\cite{BR,ME}.
Surprisingly, this permits to study \(\mathcal Hom_{\mathcal A}(M)\) where 
\(\mathcal A\) is a finitely generated algebra and \(M\) has a single 
generator\cite{GR}. In particular one can obtain a certificate \(cert(M)\) 
checking 
whether the module is or not indecomposable.
Exploiting the degrees of freedom in the intermediate computations, one can 
study in  complete detail the moduli of decompositions of \(M\).
Applications can be designed in every characteristic \((c)\). Here are given
 two of them:
\begin{itemize}
\item decomposition of boolean functions \((c=2)\). This provides a criterion of 
complexity usable in cryptography.
\item decomposition of combinatorial modules \((c=0)\).
\end{itemize}

This new method is intended to take place in MuPAD-Combinat.

\bigskip
\noindent
{\bf Example:}

\begin{figure}
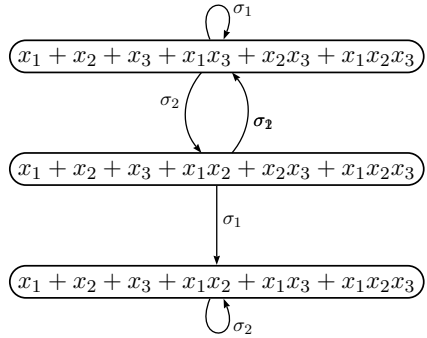

\begin{center}
\SmallPicture\VCDraw{
\begin{VCPicture}{(0,-5)(8,3)}
\ChgEdgeLabelScale{0.8}
\StateVar[x_1+x_2+x_3+x_1x_3+x_2x_3+x_1x_2x_3]{(3,4)}{A}
\StateVar[x_1+x_2+x_3+x_1x_2+x_2x_3+x_1x_2x_3]{(3,1)}{B}
\StateVar[x_1+x_2+x_3+x_1x_2+x_1x_3+x_1x_2x_3]{(3,-2)}{C}
\ArcR{A}{B}{\sigma_2}
\EdgeL{B}{C}{\sigma_1}
\ArcR{B}{A}{\sigma_2}
\ArcR{D}{B}{\sigma_1}
\CLoopN[.75]{A}{\sigma_1}
\CLoopS[.75]{C}{\sigma_2}
\end{VCPicture}}
\end{center}
\caption{
Action of \(\sigma_i\) on  
\(
g= \Z/2\Z[S_3] (x_1+x_2+x_3+x_1x_3+x_2x_3+x_1x_2x_3)
\)
}\label{f1}
\end{figure}

\begin{figure}
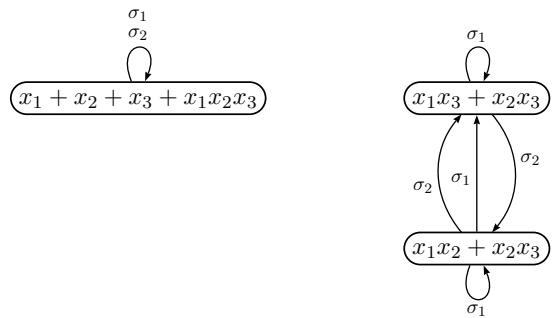

\begin{center}
\SmallPicture\VCDraw{
\begin{VCPicture}{(0,10)(8,3)}
\ChgEdgeLabelScale{0.8}
\StateVar[x_1+x_2+x_3+x_1x_2x_3]{(-1,8)}{F}
\StateVar[x_1x_3+x_2x_3]{(8,8)}{A}
\StateVar[x_1x_2+x_2x_3]{(8,04)}{B}
\ArcL{A}{B}{\sigma_2}
\ArcL{B}{A}{\sigma_2}
\CLoopN[.5]{F}{\StackTwoLabels{\sigma_1}{\sigma_2}}
\CLoopN[.5]{A}{\sigma_1}
\EdgeL{B}{A}{\sigma_1}
\CLoopS[.5]{B}{\sigma_1}
\end{VCPicture}}
\end{center}
\caption{Complete maximal decomposition of \(M=g\cdot \mathcal F\)}\label{f2}
\end{figure}

\noindent
We consider the boolean function \(f:\{0,1\}^3\rightarrow \{0,1\}\) 
defined by \(f(x_1,x_2,x_3)=x_1x_2+x_1+x_3\), the action being given 
by the algebra of the symmetric group permuting the variables
(\(\mathcal A= \Z/2\Z[S_3]\)).
 We apply our algorithm on this
 function and obtain figure \ref{f1}  which represent a complete maximal 
decompostion of the module. When we apply the algorithm, we deduce that  the 
 module \(M\) can be decomposed into \(M_1\oplus M_2\) with 
\(M_1={\Z}/2{\Z}[S_3](x_1x_2+x_1x_3+x_2x_3)\) and 
\(M_2={\Z}/2{\Z}[S_3](x_1+x_3+x_1x_2+x_2x_3)\)
(see figure \ref{f2}).
\newpage


%
%

\end{document}